# USING UPGRADED VERSIONS OF CLOSE APPROACH MANEUVERS AS TRANSPORTATION SOLUTIONS FOR DEEP SPACE MISSIONS

Antonio F. Bertachini A. Prado

Instituto Nacional de Pesquisas Espaciais – INPE

antonio.prado@inpe.br

**Abstract**

Gravity-Assisted maneuvers have been used as a technique to reduce fuel consumption in deep space missions for several decades now. It opened the doors of the exterior solar system. The literature shows those results, as well as new versions of this maneuver, which includes: the use of propulsion combined with the close approach, both high or low thrust; the passage by the atmosphere of a planet to help to change the trajectory of the spacecraft; the use of tethers to increase the changes in the velocity of the spacecraft. All those new versions have the goal of increasing the variations of energy given by the maneuver, making possible missions that would not be possible without this technique. In particular, the use of tethers is theoretical very promising, both in terms of giving extra energy to the spacecraft as well as making possible the use of smaller bodies for the closest approach. It gives much more flexibility to mission designers, which can plan missions using a large variety of smaller celestial bodies such as asteroids and even comets as a final goal of the mission or an intermediate step to observe the body and to get extra energy for the spacecraft. The idea of the construction of an "Escape Portal" and a "Capture Portal" using tethers has also been discussed in the literature, showing large gains in terms of energy for a spacecraft. The idea is to use tethers based maneuvers to send the spacecraft to the exterior solar system and also to capture that spacecraft after reaching its destination.

The construction of this portal would allow an unlimited number of maneuvers using the same tether and celestial body, which would be very beneficial for deep space missions using small satellites. This structure would be formed by a tether that remains fixed in an asteroid. At the other end of the tether, a large net is fixed, such that the only action required by the spacecraft to make the maneuver is to hit the net.

In that sense, the main goal of the present research is to combine those two ideas to allow a complete mission based on tethers that can use one tether to send the spacecraft to the exterior of the solar system and another one to capture it. It would make a complete mission that does not need fuel, at least for the most expensive part of the maneuver. Only secondary maneuvers would use propulsion. This idea needs to be better developed in the future, but can open the possibility of missions that are not possible with the current technology if made based in propulsion systems.

## 1. Introduction

Deep space missions usually have large fuel consumption and this is the most important limitation in many situations, so minimizing this factor is crucial to make more missions possible. Those costs are large when sending the spacecraft from Earth, as well as when making the capture in the target body. Other orbital maneuvers may be required, but they have much smaller fuel consumptions. In many missions, the spacecraft makes just a passage by the target body, which reduces the scientific return of the mission, because the fuel required for orbit insertion is impossible to be achieved. Several techniques have been discussed to solve this problem, including orbital maneuvers, gravitational capture and gravity-assisted maneuvers.

The gravity-assisted maneuver is the most used one and is explained in Broucke [1], Dowling et al [2] and Carvell [3]. Combinations with propulsion are also considered, as shown in Casalino et al. [4] and Petropoulos et al. [5]. A very famous mission that used this technique is the Voyager, which is described in Minovich [6] and Kohlhase and Penzo [7].

The size and masses of celestial bodies, in particular the moons of the Solar System and asteroids, impose a limitation in the energy gains given by swing-bys. An interesting solution is the use of tethers to make the rotation of the spacecraft [8-14].

Following this idea, the rotation of the spacecraft is based on the tether and can be made around asteroids or any other small body of the solar system. The literature also shows how to use this technique to help to send a spacecraft to deep space [14] and how to use planetary moons to capture a spacecraft coming in high velocity [13]. As a sequence of this research, the present paper proposes a combination of an "Escape

Portal" and a "Capture Portal", such that a complete mission can be made using tethers, to give energy for a spacecraft to go to deep space and then using another tether to capture the spacecraft using a planetary moon or an asteroid that is far from the Sun.

The structure of both portals is based in a tether fixed in the asteroid or a planetary moon and the other end have a large net, so the spacecraft can make the maneuver. Figure 1 shows a scheme for this system for the "escape portal", but the idea is similar for the "capture portal" [14].

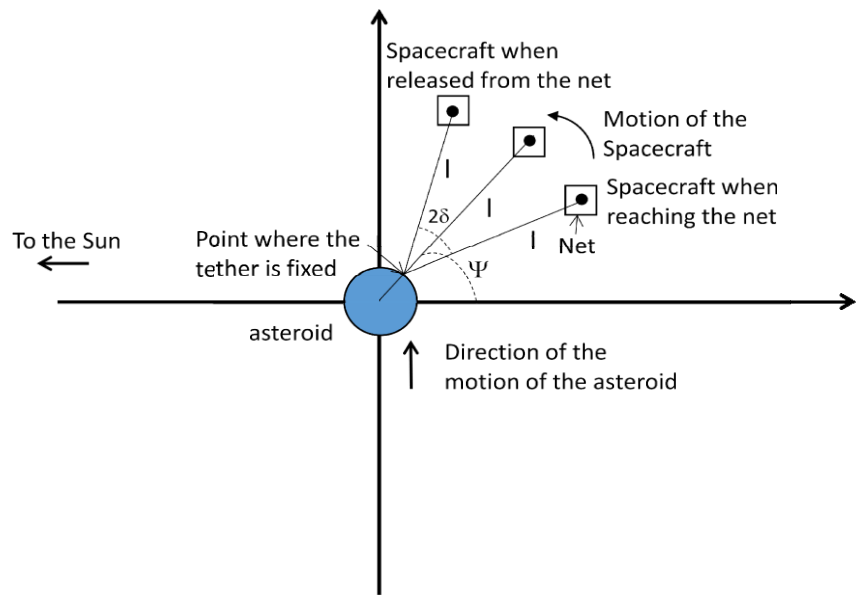

**Fig. 1. The "Escape Portal".**

The mathematical model used is the restricted planar elliptical three-body problem [15] when the tether is not active, because most of the asteroids are in elliptical orbits around the Sun. When the spacecraft is attached to the asteroid, we consider that the velocity vector rotates by a given angle, with a constant angular velocity. This maneuver does not depend on the length of the tether and the mass of the asteroid [14].

Therefore, we have a new potential form to give large amounts of energy to the spacecraft to help it to go to the exterior Solar System, without using propelled maneuvers. To make real applications, it is necessary more detailed studies, like considering the effects of the mass of the tether; the modification of the orbit of the asteroid due to the maneuver; the construction of strong and light tethers; as well as to develop accurate techniques to release the spacecraft from the net, as well as other practical details.

## 2. Definition of the Problem and Mathematical Models

The dynamical system has three bodies that are assumed to be points of mass: the asteroid, the Sun and the spacecraft, which has a negligible mass. The asteroid is orbiting the Sun in an elliptical orbit, so we have an elliptical restricted three-body problem [15].

The objective is the verification of the two-body energy and angular momentum of the spacecraft with respect to the Sun before and after the maneuver, so we can detect escapes and captures.

To perform the escape maneuver, the spacecraft approaches the asteroid, hit the net fixed in the tether, makes the rotation around the asteroid and then leaves the net. It is assumed that the tether is not affected by the maneuver.

The variables used to identify the trajectories are: 1) $V_{inf}$, the velocity of the spacecraft with respect to the asteroid when it approaches for the passage; 2) $\psi$, the angle between the point that specifies the middle of the trajectory around the asteroid and the line Sun-asteroid; 3) $l$, the constant size of the tether, 4) $\delta$, half of the angle that the velocity vector of the spacecraft rotates around the asteroid.

The spacecraft moves under the equations given by the planar elliptical restricted three-body problem [15] in the fixed system of reference with the origin in the center of mass of the two primaries (Sun and asteroid). The horizontal axis x is the line connecting the Sun and the asteroid and the vertical axis y is perpendicular to x. In this system, the Sun and the asteroid are in elliptic orbits around the center of mass, with positions given by: $x_1 = -\mu r_s \cos v_S$, $y_1 = -\mu r_s \sin v_S$, $x_2 = (1-\mu) r_a \cos v_a$, $y_2 = (1-\mu) r_a \sin v_a$, where (x$_1$, y$_1$) are the coordinates of the Sun, (x$_2$, y$_2$) are the coordinates of the asteroid, r$_S$ is the distance between the Sun and the origin of the system, given by the expression $r_S = (1-e^2)/(1+e\cos v_S)$, with $v_S$ the true anomaly of the Sun; $r_a$ is the distance between the asteroid and the origin of the system, given by the expression $r_a = (1-e^2)/(1+e\cos v_a)$, with $v_a$ the true anomaly of the asteroid. The equations of motion are shown in Eqs. (1-2).

$$\ddot{x} = \frac{-(1-\mu)(x-x_1)}{r_1^3} - \frac{\mu(x-x_2)}{r_2^3} \qquad (1)$$

$$\ddot{y} = \frac{-(1-\mu)(y-y_1)}{r_1^3} - \frac{\mu(y-y_2)}{r_2^3} \qquad (2)$$

where the double over dots represent the second derivative with respect to time; $r_1$ and $r_2$ are the distances spacecraft-Sun and spacecraft-asteroid, respectively, given by the expressions $r_1^2 = (x - x_1)^2 + (y - y_1)^2$, $r_2^2 = (x - x_2)^2 + (y - y_2)^2$; x and y are the coordinates of the spacecraft; μ is the mass parameter of the system (the ratio between the mass of the smaller primary and the total mass of the system). This canonical system of units uses the semi-major axis of the two primaries as the unit of distance; the total mass of the system as the unit of mass and the unit of time is defined such that the period of the motion of the two primaries is 2π.

The numerical algorithm involves integrating the equations of motion of the spacecraft in positive times first, so the spacecraft leaves the "Escape Portal". Then it travels under the elliptical restricted three-body problem dynamics, with initial conditions given at the point where the spacecraft leaves the "Escape Portal". The integration is performed until the spacecraft reaches a distance of 0.5 canonical units. At this point the two-body energy and angular momentum Sun-spacecraft can be measured and considered constant after this time. Then, the numerical integration is performed in negative times, starting at the point where the spacecraft reaches the "Escape Portal". This numerical integration is also performed until the spacecraft reaches a distance of 0.5 canonical units from the asteroid. Then, the energy and angular momentum of the asteroid with respect to the Sun are computed. They are the values for the orbit before the close approach with the asteroid. Considering the four possibilities for the orbit before the maneuver (open direct, open retrograde, closed direct and closed retrograde) and the same possibilities for the orbit after the maneuver, it is possible to define sixteen classes of trajectories. In this way, it is possible to show the results of the passages by the portal in letter-plots that specify what happened to the orbit due to the close approach, both in terms of energy and direction of the orbit. A particular attention is given to orbits that result in escapes from the system, because they are the ones that insert the spacecraft into an open orbit around the Sun, so leaving the Solar System without the use of propulsive forces.

Therefore, the algorithm has the following steps.

i) Values are given for the initial conditions *l*, $\psi$, Vinf, $\delta$;

ii) The initial state, when the spacecraft leaves the "Escape Portal", assuming that the tether is perpendicular to the velocity of the spacecraft to minimize the tension [8], are: $X_{i1} = l\cos(\nu + \psi + \delta) + (1 - \mu)r_{sa}\cos(\nu)$, $Y_{i1} = l\sin(\nu + \psi + \delta) + (1 - \mu)r_{sa}\sin(\nu)$, $V_{xi1} = V_{ra}\cos(\nu) - V_{ta}\sin(\nu) - V_{inf}\sin(\nu + \psi + \delta)$, $V_{yi1} = V_{ra}\sin(\nu) + V_{ta}\cos(\nu) + V_{inf}\cos(\nu + \psi + \delta)$, where $V_{ra}$ and $V_{ta}$ are the radial and tangential velocities of the asteroid with respect to the Sun, respectively; $r_{sa}$ is the asteroid-Sun distance; $\nu$ is the true anomaly of the asteroid at the moment of the maneuver; μ is the non-dimensional mass of the asteroid;

iii) The equations of motion are then integrated forward in time and the two-body energy ($E_+$) and angular momentum ($C_+$) of the spacecraft with respect to the Sun after the maneuver are calculated;

iv) The equations of motion are then integrated backward in time, starting when the spacecraft reaches the net. The initial state is: $X_{i2} = l\cos(\nu + \psi - \delta) + (1 - \mu)r_{sa}\cos(\nu)$, $Y_{i2} = l\sin(\nu + \psi - \delta) + (1 - \mu)r_{sa}\sin(\nu)$, $V_{xi2} = V_{ra}\cos(\nu) - V_{ta}\sin(\nu) - V_{inf}\sin(\nu + \psi - \delta)$, $V_{yi2} = V_{ra}\sin(\nu) + V_{ta}\cos(\nu) + V_{inf}\cos(\nu + \psi - \delta)$, and then the two-body energy ($E_-$) and angular momentum ($C_-$) Sun-spacecraft before the maneuver are measured;

v) It is then possible to observe if a capture ($E_- > 0$ and $E_+ < 0$) or an escape ($E_- < 0$ and $E_+ > 0$) occurs, as well as the modifications in the sense of the orbits.

## 3. Results

The results consist of graphs showing the effects of the portals in the trajectories of the spacecraft, including regions of captures and escapes. Those plots show the regions of the angle and velocity of approach that resulted in escapes and captures of the spacecraft with respect to the Sun. So, by calculating the typical values of the approach velocity for each particular system, it is possible to know if the energy given or removed by the portal is enough or not to generate escapes or captures. As a first example, it is showed the escape maneuvers using a fictitious asteroid with semi-major axis of 400,000,000 km, eccentricity of 0.8, zero inclination and mass parameter of $10^{-10}$. The cable has length of l = 20 km and it is assumed a maximum rotation, which means that $\phi$ = 90°. The choice of the asteroid has to follow the rules: i) it has to be small enough, so the tether can be small and still avoid a collision between the spacecraft and the asteroid; ii) the velocity around the Sun should be large, because it gives more energy to the spacecraft; iii) its velocity of approach should be inside the range of escapes from the Solar System, based in the letter-plots made in the present paper.

Table 1 shows the definition of the classes of orbits according to the modifications obtained in the trajectory of the spacecraft, in terms of being open/closed or direct/retrograde. The letter "Z" means that the spacecraft did not reach the distance limit of 0.5 canonical units from the asteroid during the integration time and the spacecraft remained captured around the asteroid. The plots have the angle of approach (in degrees) on the horizontal axis, the velocity of approach (in km/s) on the vertical axis and a letter assigned to each point to represent what happened to the trajectory of the spacecraft, as done in Broucke [1] and Prado [13].

**Table 1** - Definition of trajectories

| Before: | After: | Direct Ellipse | Retrograde Ellipse | Direct Hyperbola | Retrograde Hyperbola |
|---|---|---|---|---|---|
| Direct Ellipse | | A | E | I | M |
| Retrograde Ellipse | | B | F | J | N |
| Direct Hyperbola | | C | G | K | O |
| Retrograde Hyperbola | | D | H | L | P |

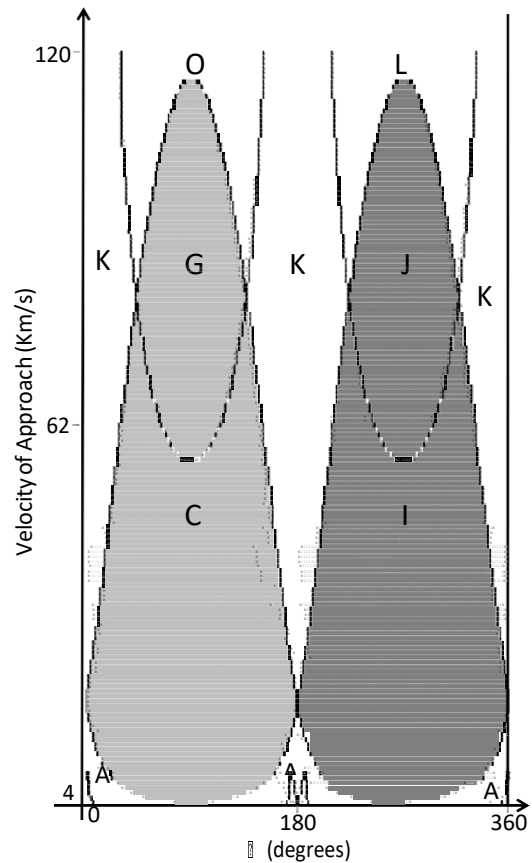

**Figure 2 – Results for the "Escape Portal" using an asteroid with a semi-major axis of 400,000,000 km, e = 0.8 and µ = 10$^{-10}$ using a cable with *l* = 20 km and making the maximum rotation ⏁ = 90 degrees.**

Figure 2 shows the results for the already cited asteroid. It is visible that captures and escapes start at velocities of approach near 4 km/s and go up to 114 km/s. There are escapes and captures for almost all values of the angle of approach.

A description of the orbits are: a) Escape orbits, letters I, J, M, N, which appear in the interval 180-360 degrees, the region of gains of energy; b) Capture orbits, letters C, D, G, H, which appear in the interval 0-180 degrees, the region with loss of energy; c) Elliptic orbits before and after the maneuver, letters A, B, E, F, which appear in the regions of smaller velocities of approach, at the bottom of the plots; d) Hyperbolic orbits before and after the maneuver, letters K, L, O, P, which appear in the regions of larger velocities of approach, in the upper parts of the plots; e) Orbits that reverse the sense of motion from direct to retrograde, letters E, M, G, O, which appear in the left part of the figures, in the interval 0-180 degrees; f) Orbits changing from retrograde to direct, letters B, D, J, L, concentrated in the interval 180-360 degrees; g) Retrograde orbits, letters F, H, N, P, which are concentrated in the middle part of the figures; h) Direct orbits, letters A, C, I, K, concentrated near the borders of the figures. The borders are also important, because they represent parabolic orbits, like A-I, B-J, F-N, H-P, I-K, J-L, N-P, N-L, F-H. There are also orbits with zero angular momentum, which means rectilinear orbitse, which are F-B, N-J, L-P, K-L, I-J, A-B.

There is a symmetry in the plots with respect to the middle vertical line, because the orbit with angle of approach ⍺ differs from an orbit with angle of approach ⍺ + 180º by a time reversal. Therefore, there are some equivalences: I-C, J-G, L-O, B-E, N-H, M-D. A, F, K and P remains unchanged. The most important ones for the present research are captures (light gray) and escapes (dark gray).

To prepare for the capture maneuver, it is necessary to obtain the energy of the spacecraft with respect to the Sun after the escape maneuver. All the information required are available, since we know the velocity and position of the spacecraft with respect to the Sun after the maneuver. Figure 3 shows the results.

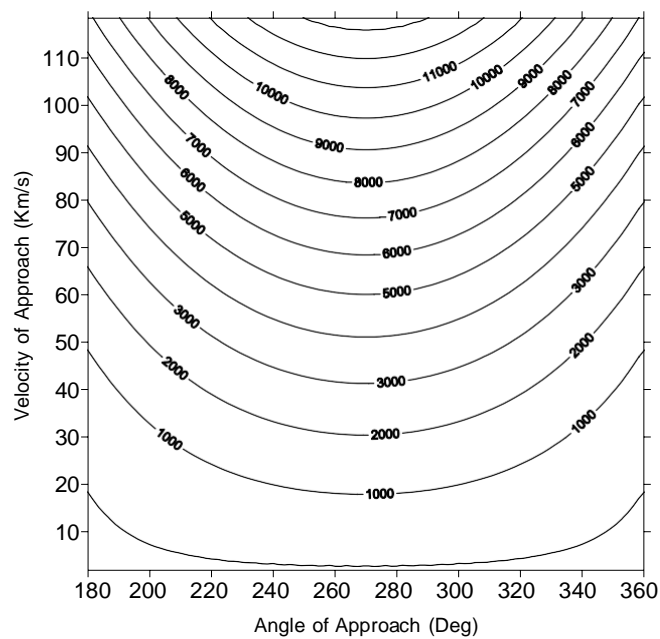

**Figure 3 - Energy of the spacecraft after passing by an asteroid with a semi-major axis of 400,000,000 km, e = 0.8 and μ = $10^{-10}$ assuming *l* = 20 km and ⍺ = 90º.**

**The capture phase**

The next step of this complete maneuver based in tethers is to obtain the range of velocity of approaches when the spacecraft reach the target. Based on the two-body energy (vis-viva equation), we can write an equation for the velocity of the spacecraft as a function of its distance from the Sun, which is the distance from the Sun to the celestial body to be visited. Equation (3) shows this result, where v is the velocity of the spacecraft with respect to the Sun, E is the two-body energy Sun-spacecraft, $\mu$ is the gravitational parameter of the Sun ($1.3275 \times 10^{20}$ m$^3$/s$^2$) and r the Sun-spacecraft distance.

$$v = \sqrt{2E + \frac{2\mu}{r}} \qquad (3)$$

As a first example, it is assumed that it is desired to orbit Jupiter and that its moon Adrastea is used for the capture. In this example, r = 484.000.000 km. Then, using Eq. (3) and Fig. 3, the velocities will be in the range 39-112 km/s with respect to the Sun. Combining with the velocity of Jupiter (13 km/s), the relative velocity of the spacecraft with respect to Jupiter will be in the range 26-125 km/s.

After those calculations, we made plots to map orbits using Adrastea for the capture, in a very similar form that was done for the escape part of the mission. The results are shown in Fig. 4. There are two plots, because we used two values for the rotation angle: 60 and 90 degrees. The length of the tether was l = 100 km. In this plot we added colors to identify better the orbits. Capture orbits are marked in light blue, red and gray. The results show clearly that captures are possible in the interval from 12 km/s to 75 km/s, so covering a large interval of velocities coming from the interior Solar System. It means that the complete maneuver is possible based in tethers is possible in a large portion of velocities of coming from the interior Solar System, from 26 to 75 km/s. It is just a question of finding the proper geometries and velocity of approach to the asteroid. Of course this concept can be generalized to other planets and moons of the Solar System.

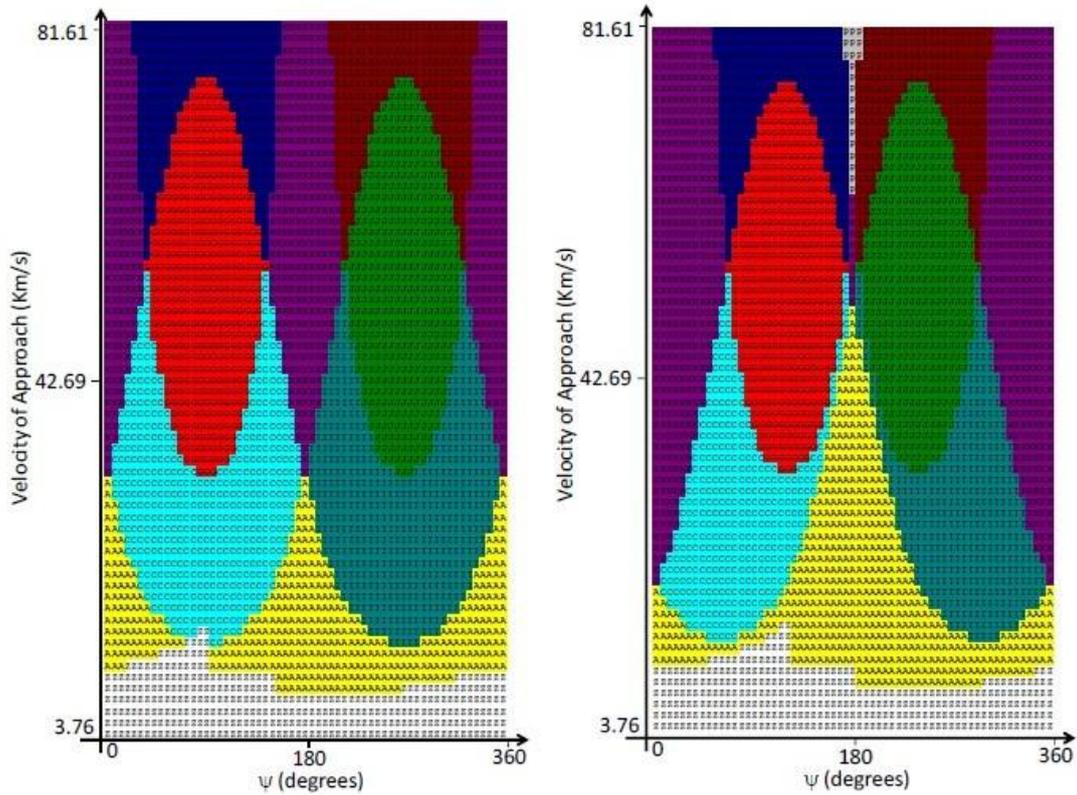

**Figure 4 - Results using Adrastrea to obtain captures around Jupiter for ≬ = 90≬ (left) and ≬ = 60≬ (right).**

### 4. Conclusions

The present paper combined the concepts of "Escape Portal" and "Capture Portal" to study a complete maneuver to send a spacecraft from the Earth to the exterior Solar System. The main idea is to use a tether to make a rotation in the spacecraft around an asteroid to give energy to the spacecraft, as done in a Swing-By maneuver, to send it far from the interior Solar System. After that, a similar idea is used to make a tethered rotation in one of the moons of the planet visited to make the capture.

Jupiter and its moon Adrastea were used as an example. The results showed that velocities of approach as small as 2 km/s can generate escapes from the Solar System, to reach Jupiter much faster and without using propulsion, then made transfers like Hohmann. The spacecraft will leave the asteroid with two-body energy Sun-spacecraft in the range from 1000 to 12000 km$^2$/s, which implies in relative velocities of the spacecraft with respect to Jupiter in the range 26-125 km/s. The simulations showed that a complete mission is possible in the interval from 26 to 75 km/s, if adequate geometries and velocity of approach to the asteroid are found.

Those ideas can be used for other asteroids, planets and moons of the Solar System and can be a powerful technique for faster and cheaper missions to the exterior Solar System.

## Acknowledgments

The author thanks the grant # 2016/24561-0, from São Paulo Research Foundation (FAPESP).

## References

1. Broucke, R. A., "The Celestial Mechanics of Gravity Assist." AIAA Paper 88-4220, 1988.

2. Dowling, R. L., Kosmann, W. J.; Minovitch, M. A.; Ridenoure, R. W., Gravity Propulsion Research at UCLA and JPL, 1962-1964. In: 41$^{st}$ Con. of the IAF, Dresden, GDR, 6-12 Oct. 1991.

3. Carvell, R. "Ulysses -the sun from above and below." *Space*, vol. 1, 1985, p. 18-55.

4. Casalino, L., Colasurdo, G. and Pasttrone D. "Optimal low-thrust escape trajectories using gravity assist." *Journal of Guidance, Control and Dynamics*, vol. 22, No 5, 1999, pp. 637-642.

5. Petropoulos, A. E., Longuski, J. M., and Vinh, N. X., Shape-based analytic representations of low-thrust trajectories for gravity-assist applications. American Astronautical Society, AAS Paper 99-337, Aug. 1999.

6. Minovich, M. A., A method for determining interplanetary free-fall reconnaissance trajectories. Pasadena: JPL , Aug. 23, 47 p. (JPL Tec. Memo 312-130), 1961.

7. Kohlhase, C. E.; Penzo, P. A., "Voyager mission description." *Space Science Reviews*. vol. 21, No. 2, 1977, pp. 77-101.

8. Lanoix, E. L. M., Misra, A. K., "Near-Earth Asteroid Missions Using Tether Sling Shot Assist." *Journal of Spacecraft and Rockets*, vol. 37, No 4, 2000, pp. 475-480.

9. Penzo, P. A.; Mayer, H. L., "Tethers and Asteroids for Artificial Gravity assist in the Solar System." *Journal of Spacecraft and Rockets*, vol. 23, No 1, 1986, pp. 79-82.

10. Lanoix, E. L. M., "Tether Sling Shot Assists: A Novel Approach to Travelling in the Solar System." *Proceedings of the 9th Canadian Aeronautics and Space Institute Conference on Astronautics*, Ottawa, Canada, 1996, pp. 62-71.

11. Puig-Suari, J., Longuski J. M., Tragesser, S. G., "A Tether Sling for Lunar and Interplanetary Exploration." *Acta Astronautica*, vol. 36, No 6, 1995, pp. 291-295.

12. Thompson, W. B., Stern, M. O., "A Skyhook from Phobos to Mars." *Proceedings of the 4th International Conference on Tethers in Space*, NASA, Washigton, DC, 1995, pp. 1737-1745.

13. Prado, A.F.B.A., Using Tethered Gravity Assisted Maneuvers for Planetary Capture, *Journal of Guidance, Control and Dynamics*, Vol. 38, 2015, pp. 1852- 1856.

14. Prado, AFBA. Tethered Gravity Assisted Maneuvers in Close Approach Asteroids to Accelerate a Spacecraft. Advances in the Astronautical Sciences, v. 156, p. 3853-3872, 2016.

15. Szebehely, V.G., Theory of Orbits, Academic Press, New York, 1967.